\documentclass[12pt]{article}
\usepackage{hyperref}
\usepackage{amsmath,amssymb}
\usepackage{url,color}
\usepackage{graphicx}
\usepackage{slashed}
\usepackage{listings} % Code
\usepackage{multirow} % more control on table columns/rows
\usepackage{booktabs} % for better looking tables
\usepackage{pifont} % checkmarks and crosses
\makeatletter
\newif\if@preliminary
\@preliminaryfalse
\def\preliminary{\@preliminarytrue}

\def\preprintno#1{\def\@preprintno{#1}}
\def\address#1{\def\@address{#1}}
\def\email#1#2{\thanks{\tt #1@{}#2}}
\def\abstract#1{\def\@abstract{#1}}
\renewcommand\abstractname{ABSTRACT}
\newlength\preprintnoskip
\newlength\abstractwidth
\@titlepagetrue
\renewcommand\maketitle{\begin{titlepage}%
  \let\footnotesize\small
  \hfill\parbox{\preprintnoskip}{%
  \begin{flushright}\@preprintno\end{flushright}}\hspace*{1cm}
  \vskip 60\p@
  \begin{center}%
    {\Large\bf\boldmath \@title \par}\vskip 1cm%
    {\sc\@author \par}\vskip 3mm%
    {\@address \par}%
    \if@preliminary
      \vskip 2cm {\large\sf PRELIMINARY DRAFT \par \@date}%
    \fi
  \end{center}\par
  \@thanks
  \vfill
  \begin{center}%
    \parbox{\abstractwidth}{\centerline{\abstractname}%
    \vskip 3mm%
    \@abstract}
  \end{center}
  \end{titlepage}%
  \setcounter{footnote}{0}%
  \let\thanks\relax\let\maketitle\relax
  \gdef\@thanks{}\gdef\@author{}\gdef\@address{}%
  \gdef\@title{}\gdef\@abstract{}\gdef\@preprintno{}
}%
\def\@citex[#1]#2{\if@filesw\immediate\write\@auxout{\string\citation{#2}}\fi
  \def\@citea{}\@cite{\@for\@citeb:=#2\do
    {\@citea\def\@citea{,\penalty\@m}\@ifundefined
       {b@\@citeb}{{\bf ?}\@warning
       {Citation `\@citeb' on page \thepage \space undefined}}%
\hbox{\csname b@\@citeb\endcsname}}}{#1}}
\def\citerange{\@ifnextchar [{\@tempswatrue\@citexr}{\@tempswafalse\@citexr[]}}
\def\@citexr[#1]#2{\if@filesw\immediate\write\@auxout{\string\citation{#2}}\fi
  \def\@citea{}\@cite{\@for\@citeb:=#2\do
    {\@citea\def\@citea{--\penalty\@m}\@ifundefined
       {b@\@citeb}{{\bf ?}\@warning
       {Citation `\@citeb' on page \thepage \space undefined}}%
\hbox{\csname b@\@citeb\endcsname}}}{#1}}
\long\def\@makecaption#1#2{%
  \vskip\abovecaptionskip
  \sbox\@tempboxa{#1: \emph{#2}}%
  \ifdim \wd\@tempboxa >\hsize
    #1: \emph{#2}\par
  \else
    \hbox to\hsize{\hfil\box\@tempboxa\hfil}%
  \fi
  \vskip\belowcaptionskip}
\makeatother
\newcommand{\GeV}{\text{GeV}}

\newcommand{\bss}{\begin{tiny}}
\newcommand{\ess}{\end{tiny}}
\newcommand{\whizard}{\texttt{WHIZARD}}
\newcommand{\oMega}{\texttt{O'Mega}}
\newcommand{\circeone}{\texttt{CIRCE1}}
\newcommand{\circetwo}{\texttt{CIRCE2}}
\newcommand{\vamp}{\texttt{VAMP}}

\newcommand{\openloops}{\texttt{OpenLoops}}
\newcommand{\gosam}{\texttt{GoSam}}
\newcommand{\recola}{\texttt{Recola}}
\newcommand{\pythiasix}{\texttt{PYTHIA~6}}
\newcommand{\fastjet}{\texttt{FastJet}}
\newcommand{\sarah}{\texttt{SARAH}}
\newcommand{\feynrules}{\texttt{FeynRules}}
\newcommand{\vegas}{\textsc{VEGAS}}
\newcommand{\sindarin}{\textsc{Sindarin}}
\newcommand{\ufo}{\textsc{UFO}}
\begin{document}

%begin{fmffile}{wopics}
%begin{empfile}

%%%%%%%%%%%%%%%%%%%%%%%%%%%%%%%%%%%%%%%%%%%%%%%%%%%%%%%%%%%%%%%%%%%%%%%%
\preprintno{%
SI-HEP-2018-06\\
DESY 18-014}

\title{%
  New Developments in WHIZARD Version 2.6}

\author{
  W.~Kilian\email{kilian}{physik.uni-siegen.de}$^a$,
  S.~Brass\email{brass}{physik.uni-siegen.de}$^a$,
  T.~Ohl\email{Thorsten.Ohl}{physik.uni-wuerzburg.de}$^b$,
  J.~Reuter\email{juergen.reuter}{desy.de}$^c$,
  V.~Rothe\email{vincent.rothe}{desy.de}$^c$,
  P.~Stienemeier\email{pascal.stienemeier}{desy.de}$^c$,
  M.~Utsch\email{manuel.utsch}{t-online.de}$^a$
}

\address{\it%
    $^a$University of Siegen, Department of Physics, 
    D--57068 Siegen, Germany\\
    $^b$W\"urzburg University, Faculty of Physics and Astronomy, 
    D--97074 W\"urzburg, Germany\\
    $^c$DESY Theory Group,
    D--22603 Hamburg, Germany
\\[3\baselineskip]
Talk presented by W.~Kilian at the International Workshop on Future Linear
Colliders 
(LCWS2017), Strasbourg, France, 23-27 October 2017. C17-10-23.2.
}

\abstract{%
  We describe recent additions to the \whizard~2 Monte-Carlo event
  generator which improve the physics description of lepton-collider
  event samples and speed up the calculation time required for
  cross sections and event generation.
}
\maketitle

\section{The \whizard\ Monte-Carlo Generator}
\label{sec:intro}
	
The \whizard\ Monte-Carlo event generator~\cite{Kilian:2007gr,WHIZARD} is a
stand-alone program that calculates cross sections, distributions, and fully
exclusive event samples of perturbative high-energy processes at colliders.
The program handles particle-decay processes as well as scattering processes
at hadron colliders and for the planned lepton colliders such as ILC and CLIC.

The program has been used to generate simulated event samples for a wide
variety of lepton-collider studies.  New developments and improvements of the
program will allow for producing event samples with a more complete
physics description, fueling refined and extended physics and detector studies
for the ILC and CLIC collider projects.

Within \whizard, functional expressions for the elementary processes are
constructed as needed in form of source code when running the program, calling
the algebraic matrix-element generator
\oMega~\cite{Moretti:2001zz,Nejad:2014sqa} for tree-level amplitudes.  QCD is
handled in the color-flow calculational scheme~\cite{Kilian:2012pz}.  This
code is compiled and linked to the main program on the fly.  Beyond leading
order in perturbation theory, \whizard\ provides an interface to
next-to-leading order (NLO)
virtual-amplitude providers \openloops~\cite{Cascioli:2011va},
\gosam~\cite{Cullen:2014yla}, and \recola~\cite{Actis:2016mpe,Denner:2017wsf}.
NLO calculations with
\whizard~\cite{Kilian:2006cj,Robens:2008sa,Binoth:2009rv,Greiner:2011mp} have
been recently automatized~\cite{Reuter:2016qbi}, implementing
infrared-collinear subtraction via the FKS~\cite{Frixione:1995ms} subtraction
scheme.

Precision calculations for $e^+e^-$ colliders require a detailed treatment of
interactions of the incoming beams.  \whizard\ describes beam structure via
the \circeone~\cite{Ohl:1996fi} and \circetwo\ modules.  \circetwo\ is a
dedicated beam-event generator that is based on fitting detailed simulation
results for the beam-beam interactions.  The program supports any mode and
degree of beam polarization.  Furthermore, the universal part of
soft-collinear photon radiation from the initial state is accounted for by an
inclusive structure-function approach for electrons, positrons, and photons.

The effects of radiation from the final state are handled by an internal call
to \pythiasix~\cite{Sjostrand:2006za}.  The internal \pythiasix\ implementation
also provides the transformation to the fully hadronic final state.
Alternatively, events can be showered using \whizard's own analytic
shower~\cite{Kilian:2011ka}, or externally via event files.  For communicating
with further external processing of events, \whizard\ supports a comprehensive
set of event-file formats.

The \whizard\ program is optimized for handling multi-particle final states
resulting from hard processes, such as the four- to eight-fermion final states
that result from the production of the heavy resonances $W$, $Z$, $H$, and $t$
at future lepton colliders.  To this end, \whizard\ follows a multi-channel
approach with concurrent phase-space parameterizations that reflect the
singularity structure of the matrix element.  Integration over phase space is
realized by the \vamp\ module~\cite{Ohl:1998jn}, which is a multi-channel
extension of the well-known adaptive Monte-Carlo integration algorithm
\vegas.

Regarding the definition of physics models, \whizard\ contains a comprehensive
predefined library of models including the SM (Standard Model),
supersymmetry~\cite{Ohl:2002jp}, Little-Higgs models, or extra
dimensions.  Beyond these hard-coded models, it contains a model
implementation interface to the \sarah~\cite{Staub:2013tta} and
\feynrules~\cite{Christensen:2010wz,Alloul:2013bka} programs, and it supports
the \ufo~\cite{Degrande:2011ua} model-definition file format.

For providing input data to the program, steering the workflow, describing
cuts and weight factors, and for internal event analysis, \whizard\ implements
a specific scripting language, \sindarin.  The language supports
calculations 
and manipulations of events and subevents, provides an interface to the
\fastjet~\cite{Cacciari:2011ma} jet-algorithm package, and enables all I/O of
the generator.  It 
allows for conditionals, evaluation loops for parameter scans, concurrent
alternative parameter sets, and matrix-element reweighting of event samples.
The internal analysis covers free-form selection criteria and enables
histograms, tables, and plots within a uniform workflow.

\section{Issues in Standard-Model Event Generation}

Physics and detector studies for future lepton colliders rely on the
availability of validated event samples for all SM processes
that are expected to be observable.  These include resonant production processes
that incorporate two or more massive SM particles, $W$, $Z$, or $H$, or
top-quark pairs which subsequently decay into light SM particles.  For precise
predictions, the calculation has to involve complete matrix elements that
include both resonant and non-resonant contributions at a given order in the
loop expansion.  Current simulation studies generally involve matrix
elements at leading order, which in the future are to be replaced by
(virtual) higher-order matrix elements.

A large database of SM event samples has been obtained in the past using
version~1 of the \whizard\ generator.  Current and future studies employ
version~2 of \whizard.  The second version offers much greater flexibility and
convencience as an application, and is set up for improved precision and to
cover a wider range of physics models and effects.

In this note, we describe issues that have emerged in the course of this
transition to an improved and more versatile framework, their physics impact,
and technical solutions.  \whizard~2 event samples have been validated by the
ILC-CLIC generator group against
\whizard~1 event samples in those contexts where this comparison is meaningful.
We have identified three areas where improvements were required, such that the
new generation of event samples can cover a larger set of processes with
reasonable approximations and good efficiency.
\begin{enumerate}
\item
  exclusive generation of semi-hard photons, which are included in the
  initial-state radiation (ISR)
  approximation but warrant a nontrivial modification of collinear
  kinematics;
\item
  kinematic distortion of resonance shapes due to the approximations involved
  in the \pythiasix\ parton-shower algorithm;
\item
  a re-implementation of \vamp\ which now supports highly parallel integration
  and event generation within the message-passing interface (MPI)
  communication model. 
\item
  A dedicated treatment of NLO contributions and resummed threshold logarithms
  for off-shell $t\bar t$ and $t\bar t H$
  processes~\cite{Nejad:2016bci,Bach:2017ggt} is covered in a separate 
  contribution to these Proceedings.
\end{enumerate}

\section{Semi-Hard ISR Photons}

A fully inclusive ISR description is a convenient method to account for
leading-logarithmic effects of multiple photon radiation from the initial
state.  The description implemented in \whizard\ includes resummed soft
photons to all orders, as well as higher-order universal collinear
contributions.  This is sufficient, given the leading-order
approximation in the SM perturbation expansion of the hard matrix elements, to
describe the dominant effect of ISR photon emission, namely energy loss and
the resulting distortion in the shape of s-channel resonances and thresholds.
It also yields the dominant QED corrections to calculated cross sections.

However, in the context of fully exclusive event generation, there is a
non-negligible fraction of events where emitted photons are not strictly soft
or collinear but carry away a measurable amount of transversal momentum.
While this effect can safely be neglected for inclusive quantities, the
resulting $p_T$ kick on hard-process kinematics can significantly distort
event shapes and distributions of more exclusive observables.

In practice, this $p_T$ mismatch has to be taken into account for both
incoming beams simultaneously.  The approach in \whizard~1 to this problem was
to sample transverse momentum for both radiated photons independently of each
other, according to the logarithmic distribution which prevails over most of
the collinear phase space.  Regarding the hard process, this was combined to
an ad-hoc kinematics modification which did violate, to some extent, either
energy or momentum conservation.  The algorithm was applied both to
cross-section integration and to exclusive event generation.  It turns out
that the actual inaccuracies were of minor importance but there was some
uncontrolled impact on the validity of the prediction near phase-space
boundaries.
 
The implementation of transverse-momentum generation in early versions of
\whizard~2 was intended to describe a more generic chain of radiation and
on-shell projections for each beam individually.  Unfortunately, such an
approach results in more drastic energy-momentum mismatches with unphysical
results for various observables.  The discrepancy is visible in the comparison
between \whizard~1 and \whizard~2 samples, such that it should not be
tolerated for practical applications.

Therefore, version 2.6 supports a new algorithm for approximating the effect
of photon transverse momentum.  The kinematics calculation takes into account
both beams simultaneously.  The new version conserves energy and momentum
exactly.  The only (inevitable) on-shell projection is applied to the
initial partons of the hard process, which themselves are unobservable and not
part of the physical event.  Furthermore, the integration is now carried out
in the strict collinear limit where the ISR approximation is defined.  Only in
the simulation pass, the generated hard-process events are modified
individually according to the logarithmic transverse-momentum distribution.
The hard event, and any radiation originating from it, is merely boosted by
that effect, while the universal behavior of the radiated photons is correctly
described.  The approximation loses its validity for $p_T$ of the order of the
hard-process scale, where a NLO SM calculation would be required to compute
the process-dependent contributions.

\section{Resonances and Parton Shower}

In the current \whizard\ framework, events corresponding to leading-order
matrix elements are combined with the \pythiasix\ parton shower module.  The
\whizard/\oMega\ matrix elements are complete in the sense that they contain
all Feynman graphs that connect the initial state to the selected final state.
Typically they contain both resonant and non-resonant contributions.

When applying QCD radiation to a colored final state, there are two effects
that must be handled in the presence of resonances.  For instance, a resonant
$W$ boson is a colorless particle, but its $q\bar q$ decay state contains
color sources and thus initiates a parton shower.  (i) If the $p_T$ of a
radiated parton off the decay products is larger than the resonance width, the
radiation effect shifts the kinematics such that the actual matrix-element
value can deviate by a large factor.  However, the matrix element is treated
in the factorization limit and thus kept unchanged by the shower module.  It
is obvious that in this situation, the factorization assumption for the
parton-shower approximation becomes invalid.  (ii) At energy scales above the
mass of the resonance, the decaying particle may be considered as a stable,
colorless entity.  Therefore, radiation in this region is suppressed.  A full
calculation would show a destructive interference of radiation from both color
sources.

The \pythiasix\ shower module allows the programmer to mimic both effects by a
variant of the shower algorithm that starts evolution at the resonance-mass
scale, as opposed to the hard-process scale.  Momentum is distributed such
that the effective resonance mass is kept unchanged.  This addresses both
issues described above.  There is a continuum (non-resonant) contribution to
the process to which this modification would not apply, but regarding the
cross section the continuum is a higher-order effect.  In fact, the modified
algorithm has been successful for describing real LEP data.  Therefore, in the
\whizard~1 simulation for $e^+e^-$ studies, each event was assigned a
resonance history, and the shower was reorganized accordingly.

\begin{figure}[p]

\begin{center}
\includegraphics[scale=0.4]{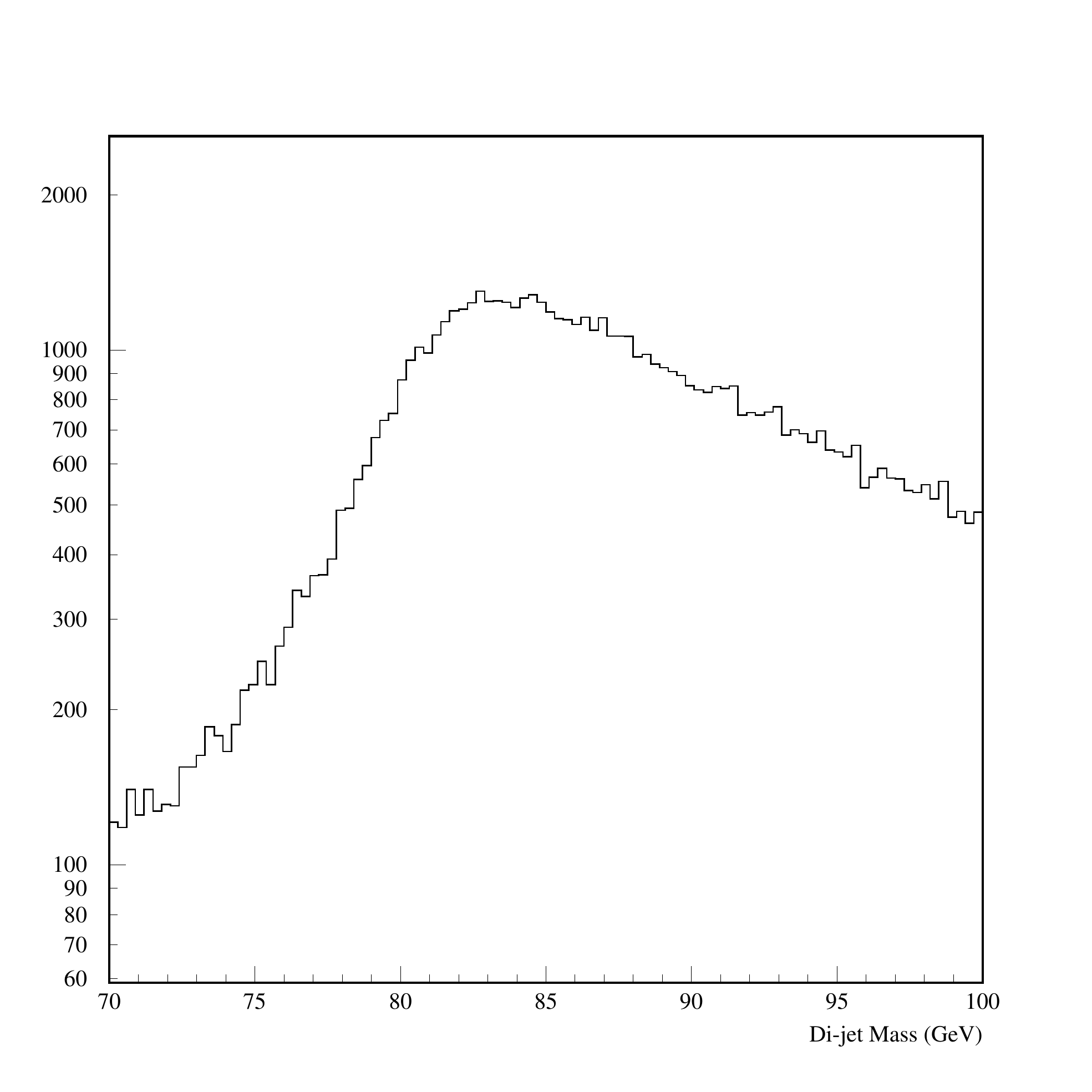}
\includegraphics[scale=0.4]{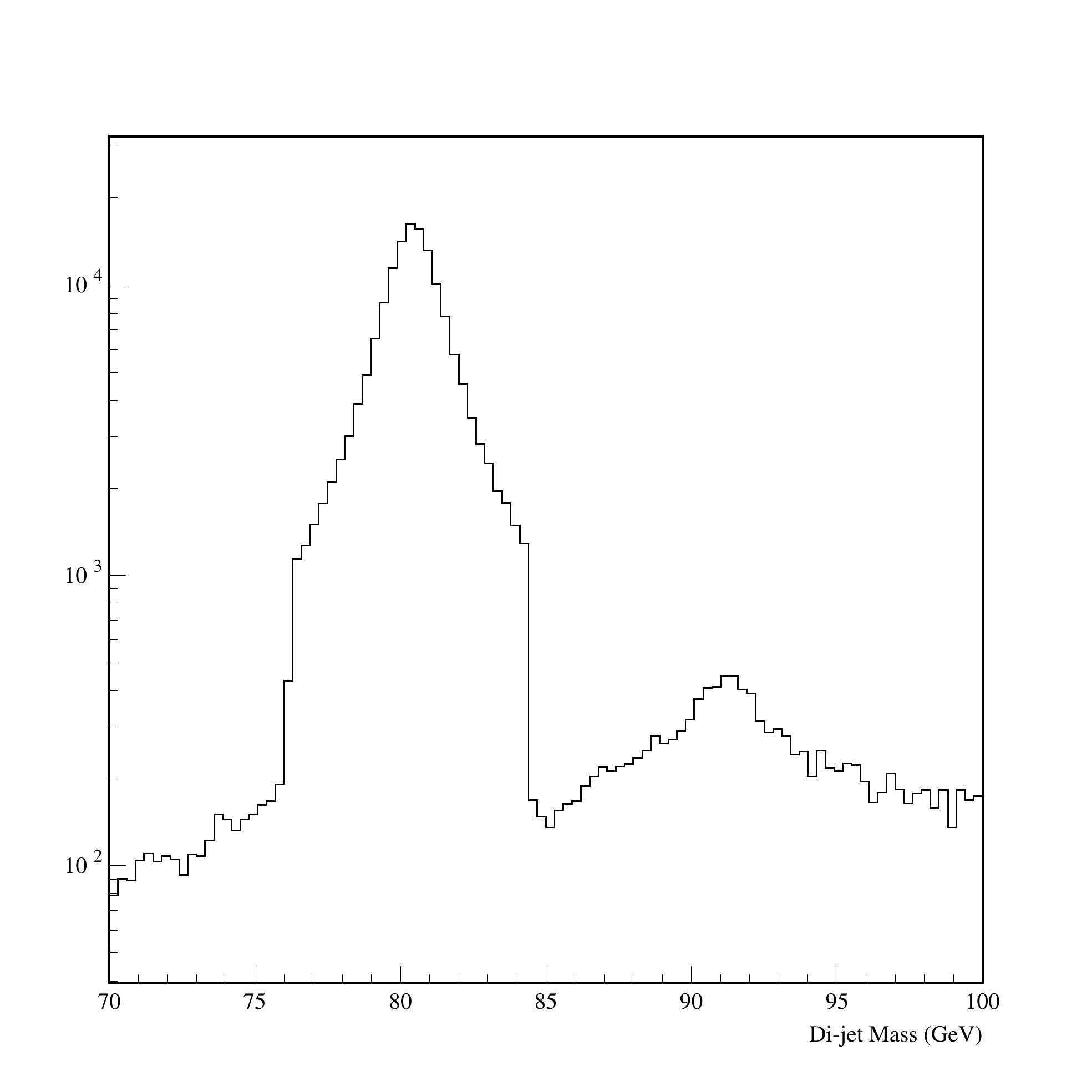}
\end{center}
\begin{center}
\includegraphics[scale=0.4]{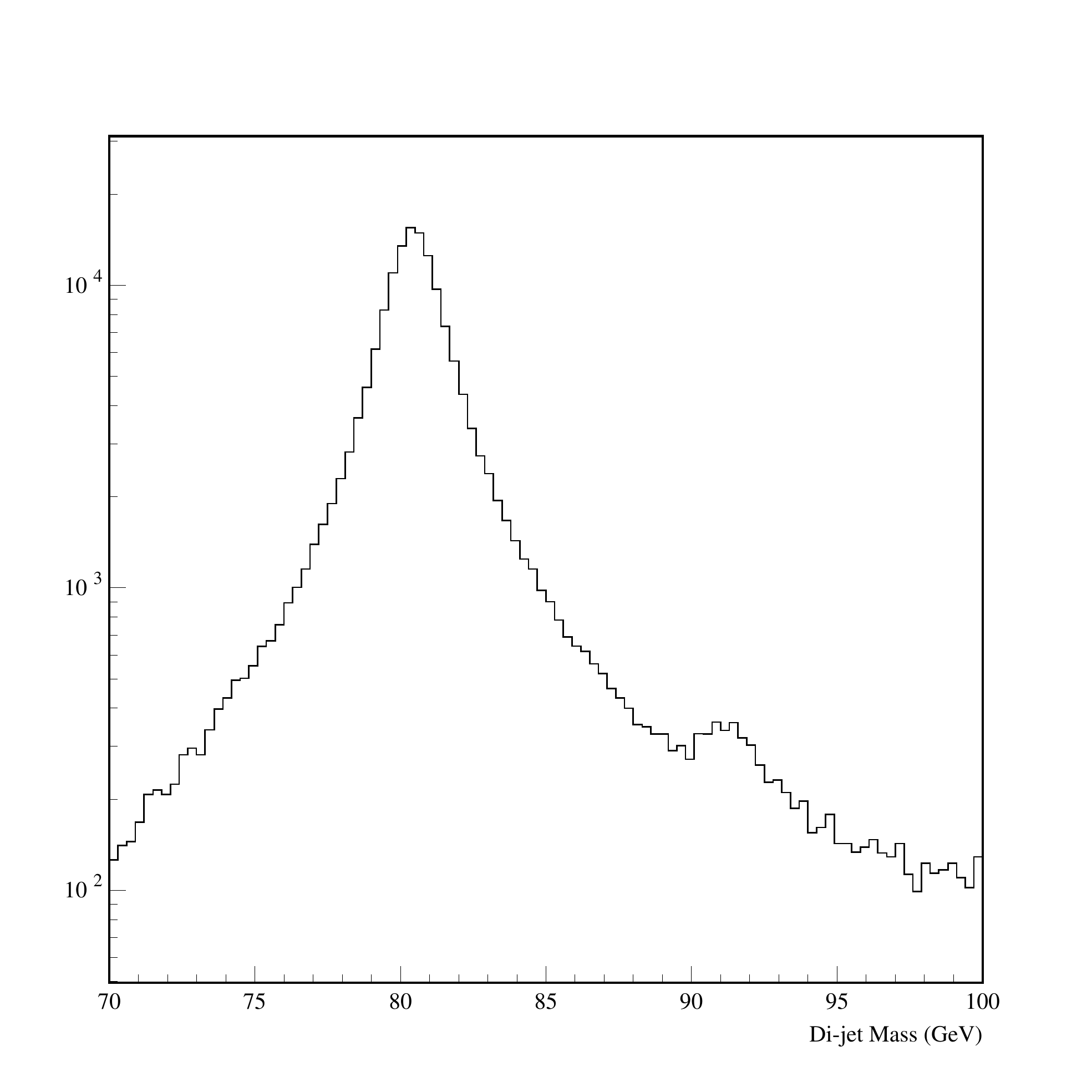}
\end{center}
\caption{Process $e^+e^-\to q\bar q q\bar q$ for $\sqrt{s}=500\;\GeV$, with
  ISR and beamstrahlung.  The plots display the dijet invariant mass
  distribution after \pythiasix\ shower for three different settings of the
  resonance-handler parameters.  Left top: no resonance history assumed; right
  top: sharp resonance-history cutoff at $(p^2-m^2)/m\Gamma = 4$; bottom:
  resonance-history cutoff at $16$, with smooth Gaussian transition factor.
  Simulation and plots by M.~Berggren.}
\label{fig:dijet}
\end{figure}

However, the accuracy of both the simulation and the expected data for ILC
call for a shower algorithm that goes beyond the leading-resonance
approximation and also holds in the kinematical regions where
neither resonance history applies, or more than one (such as $W^+W^-$/$ZZ$
with identical final state).  Moreover, any explicit NLO-QCD corrections are
not matched correctly in the simple scheme described above.  To avoid this
complication, \whizard~2 initially did not account for resonance histories at
all.  Unfortunately, this also results in unphysical resonance-shape
distortions which invalidate observables that depend on resonance shapes.

In \whizard~2.6, we have implemented an algorithm which refines the
\whizard~1 variant as adapted by the experimental lepton-collider
collaborations and is well suited to describe showering both on top of a
resonance and in the continuum, as well as in the transition region.  The
effect has been validated by explicit calculation and comparison to \whizard~1
simulations.

For each given event, the generator does not just compute the complete matrix
element, but also any possible factorized matrix element which results from
removing all graphs that do not contain a particular set of resonances.  The
resonance histories are selected according to a suitable kinematics criterion.

Matrix elements that are calculated from a restricted subset of graphs,
evaluated off resonance, depend on the chosen electroweak gauge.
Fortunately, the resulting ambiguity is controlled.  The distinction between
gauge bosons of a broken symmetry and massive vector bosons can matter only for
scales above the gauge boson masses.  Therefore, as long as the involved
off-shell distances are small compared to the corresponding resonance masses,
the ambiguity is parameterically of higher order.  In practice, the resonance
criterion measures the distance $p^2-m^2$ to each given resonance and compares
this with $m\Gamma$, where $\Gamma$ denotes the resonance width.  If this ratio exceeds a (tunable) factor, the resonance history is discarded.  For the
histories that are kinematically allowed, the factorized squared matrix
elements are related to the complete squared matrix elements.  These ratios
are interpreted as probabilities pertaining to the considered histories.  For
each event, one of the histories, or the continuum case, is selected according
to those probabilities.  Finally, the selected history is reconstructed in the
event record and thus transferred to the shower generator, which will restrict
radiation accordingly.

As a refinement, we implement a smooth transition between resonance and
continuum off shell by reweighting the factorized matrix elements with a
Gaussian.  This eliminates artefacts of the off-shell cutoff that otherwise
would show up in exclusive distributions.

In Fig.~\ref{fig:dijet}, the effect is illustrated quantitatively for the
process $e^+e^-\to q\bar q q\bar q$ which contains contributions from both
$W^+W^-$ and $ZZ$ resonance intermediate states.  If no resonance history is
imposed on the \pythiasix\ shower, the generated gluon radiation completely
washes out the resonance shapes of the $W$ and $Z$ resonance peaks at $80$ and
$91\;\GeV$, respectively.  If we insert the corresponding resonance histories
for events where the partonic dijet invariant mass is close to a resonance
peak, say for $(p^2-m^2)/(m\Gamma)<4$, shower evolution starts only at the
weak-boson mass scale, and radiation kinematics leave the $W$ and $Z$ peaks
intact.  However, due to the smallness of the non-resonant background for this
process, abruptly switching from a resonance assumption to a background
assumption at a fixed distance from the peak introduces a step in the
distribution as an unphysical artefact.  

This problem is eliminated if we implement a smooth transition from the
resonance to the background hypothesis, as shown in Fig.~\ref{fig:dijet},
bottom.  We emphasize that all variants are formally consistent with the given
order of the matrix-element calculation and of the parton shower.
Nevertheless, only a smooth transition of resonance to continuum hypothesis as
an input to the parton shower is expected to actually emulate the true
result, which otherwise would require an explicit NLO/NNLO matching
calculation.

\section{Parallel Evaluation of Adaptive Phase Space}

The running time of Monte-Carlo integration and event-generation programs can
be reduced by a significant factor if the evaluation exploits the inherent
parallelizability of evaluating a large sample of phase-space points.  The
new \vamp\ implementation in \whizard\ 2.6 aims at realizing this potential.

We have embedded the \whizard\ program, in particular the multi-channel
integration module, in a multi-processing model according to the MPI 3
standard.  The program runs on a set of computation nodes with separate
associated memory and a protocol for communicating data between nodes.

Besides actually computing matrix elements on distinct phase-space points in
parallel, the main issue is reducing the impact of communication, and limiting
the parts of the program that have to be evaluated serially.  A major
communication part, within the \vamp\ adaptive algorithm, is caused by
exchanging the grids, i.e., the sets of binning data for individual
phase-space parameterization, between nodes.  For realistic processes such as
$2\to 6$ or $2\to 8$ configurations with nontrivial helicity and color
structure, the amount of data to be exchanged can become substantial.  We are
using the MPI 3 feature of asynchronous communication to minimize the amount
of mutual blocking.

Further parallel speedup can be achieved if the multi-core architecture of
modern CPUs as computing nodes can be exploited.  To this end,
\whizard\ enables parallel evaluation of distinct helicity configurations on a
single node with multiple cores, following the OpenMP shared-memory protocol.

The MPI and OpenMP parallel features are available in the current release
version of \whizard\ and are being used for computing-intensive
studies~\cite{Anders:2018gfr}.  It turns out that OpenMP parallelization
scales well for standard multi-core processors, as expected given typical
values for helicity combinations.  The possible speedup due to multi-processor
MPI parallelization is currently exhausted up to $O(10\dots 100)$ processors,
depending on the involved process.

One of the limitations is caused by the algorithm that constructs the
phase-space parameterizations, which is inherently serial.  A
re-implementation of this algorithm, which makes use of the process structure
information which is known to the \oMega\ matrix element generator, eliminates
this problem.  A systematic study of benchmark processes and the benefit of
parallel evaluation is under way, and we expect further improvements in
upcoming \whizard\ versions.

%%%--- Bibliography ---%%%
\bibliographystyle{unsrt}

\end{document}